\documentclass[twocolumn, superscriptaddress, showpacs, amsmath, amssymb, aps, prl]{revtex4-1}

\draft 
\usepackage{amsmath}
\usepackage{xcolor}

\usepackage{graphicx}
\usepackage{hyperref}
\usepackage{placeins} 
\begin{document}

\newcommand{\blue}[1]{\textcolor{blue}{#1}}
\newcommand{\red}[1]{\textcolor{red}{#1}}


\title[Sample title]{Resonant Emittance Mixing of Flat Beams in Plasma Accelerators}

\author{S.~Diederichs}
\affiliation{Deutsches Elektronen-Synchrotron DESY, Notkestr. 85, 22607 Hamburg, Germany}
\affiliation{CERN, Espl. des Particules 1, 1211 Geneva, Switzerland}

\author{C.~Benedetti}
\affiliation{Lawrence Berkeley National Laboratory, 1 Cyclotron Rd, Berkeley, California 94720, USA}%

\author{A.~Ferran~Pousa}
\affiliation{Deutsches Elektronen-Synchrotron DESY, Notkestr. 85, 22607 Hamburg, Germany}

\author{A.~Sinn}
\affiliation{Deutsches Elektronen-Synchrotron DESY, Notkestr. 85, 22607 Hamburg, Germany}

\author{J.~Osterhoff}
\affiliation{Deutsches Elektronen-Synchrotron DESY, Notkestr. 85, 22607 Hamburg, Germany}
\affiliation{Lawrence Berkeley National Laboratory, 1 Cyclotron Rd, Berkeley, California 94720, USA}

\author{C.\,B.~Schroeder}
\affiliation{Lawrence Berkeley National Laboratory, 1 Cyclotron Rd, Berkeley, California 94720, USA}
\affiliation{Department of Nuclear Engineering, University of California, Berkeley, California 94720, USA}

\author{M.~Thévenet}
 \email{maxence.thevenet@desy.de}
\affiliation{Deutsches Elektronen-Synchrotron DESY, Notkestr. 85, 22607 Hamburg, Germany}

\date{\today}

\begin{abstract}

Linear colliders rely on high-quality flat beams to achieve the desired event rate, while avoiding potentially deleterious beamstrahlung effects.
Here, we show that flat beams in plasma accelerators can be subject to quality degradation due to emittance mixing.
This effect occurs when the beam particles' betatron oscillations in a nonlinearly coupled wakefield become resonant in the horizontal and vertical planes.
Emittance mixing can lead to a substantial decrease of the luminosity, the main quantity determining the event rate.
In some cases, the use of laser drivers or flat particle beam drivers may decrease the fraction of resonant particles and, hence, mitigate emittance deterioration.


\end{abstract}


\maketitle 


Plasma-based accelerators~\cite{Tajima:1979, Chen:1985} are promising candidates as drivers for future linear colliders due to their $\gtrsim$ GV/m accelerating gradients. Although experimental progress in terms of energy gain~\cite{Blumenfeld:2007, Gonsalves:2019, Aniculaesei:2024}, energy transfer efficiency~\cite{Litos:2014}, and energy spread preservation~\cite{Kirchen:2021, Lindstrom:2021} have increased the interest in plasma-based linear colliders~\cite{Foster:2023, Adli:2013, Schroeder:2023}, additional challenges must be overcome.

For optimal operation of a linear collider, the event rate and, consequently, the luminosity $\mathcal{L}$ must be maximized while deleterious beamstrahlung effects~\cite{Augustin:1978} must be minimized. Because the former scales as $1/(\sigma_x \sigma_y)$~\cite{Schulte:2016} (where $\sigma_x$ and $\sigma_y$ are the rms beam sizes at the interaction point in the horizontal and vertical plane, respectively) and the latter as $1/(\sigma_x + \sigma_y)$~\cite{Schroeder:2022}, a common solution is to operate with flat beams, $\sigma_x \gg \sigma_y$ (i.e., with an aspect ratio $\sigma_x / \sigma_y \gg 1$). This motivates the creation of beams with $\epsilon_x/\epsilon_y \gg 1$ (where $\epsilon_{[x,y]}$ is the beam emittance in $[x,y]$), and the preservation of this ratio during acceleration.
Established mechanisms that lead to deleterious exchange or mixing of the transverse emittances are linear coupling~\cite{Edwards:2008}, e.g., due to misaligned or skew quadrupoles, and nonlinear coupling, e.g., due to space-charge effects~\cite{Montague:1968, Hofmann:2017}, which are mainly relevant at low energies. The latter is linked to the Montague resonance that occurs if the focusing in the horizontal and vertical planes is in phase. 
Emittance mixing can occur when the motion in the $x$ and $y$ planes is coupled (for instance when the transverse force in $x$ depends on $y$), but such effects have not been previously described in plasma-based accelerators.

Plasma accelerators are often operated in the so-called blowout regime, where the driver is strong enough to expel all plasma electrons, creating a trailing ion cavity in its wake. In the ideal case of a uniform background ion distribution within the cavity, the transverse wakefields in $x$ and $y$ are decoupled, preventing emittance exchange. In practice, various nonlinear effects can perturb the transverse wakefields and cause coupling and, hence, emittance mixing. Such effects occur for collider-relevant beams that require high charge ($\sim$nC) and low emittance ($\sim100$ nm) and therefore generate extreme space-charge fields capable of ionizing the background plasma to higher levels~\cite{Bruhwiler:2003} or causing ion motion~\cite{Rosenzweig:PRL:2005, An:PRL:2017, Benedetti:PRAB:2017}, both of which can lead to the formation of nonlinearly coupled wakefields. 
Nonlinear wakefields are sometimes desired: for instance, nonlinearities in the wake due to ion motion can suppress the hosing instability~\cite{Mehrling:PRL:2018, Diederichs:2022:WBS, Diederichs:2022:DBS} while still allowing for witness beam emittance preservation through advanced matching schemes~\cite{Benedetti:PRAB:2017, Benedetti:PoP:2021}.

In this Letter, we demonstrate by means of theory and 3D particle-in-cell (PIC) simulations that coupled wakefields in plasma accelerators can lead to severe emittance mixing of flat beams when there is a resonance between the betatron oscillations in the horizontal and vertical planes for a large fraction of beam particles. With this effect, the horizontal emittance decreases as the vertical one increases, resulting in an overall growth of their geometric average and, hence, a reduction in luminosity.
This mechanism is different from nonlinearity-induced mismatch, by which a beam with a position-momentum distribution not matched to nonlinear fields relaxes at the cost of emittance growth. Unlike emittance mixing, mismatch causes emittance growth in both planes independently.
Without proper mitigation, mixing can cause a flat beam to become round, resulting in a considerable decrease in luminosity (e.g., by a factor of 50 for an initial aspect ratio of 100), while simultaneously loosing the beneficial suppression of beamstrahlung. This mechanism has direct impact on any future plasma collider design using flat beams. It has previously not been documented, since only short distance acceleration of flat beams was considered~\cite{An:PRL:2017}.

Emittance mixing for flat beams in coupled, nonlinear wakefields is illustrated with a plasma-wakefield accelerator setup in the blowout regime that resembles the first stage of the proposed HALHF collider~\cite{Foster:2023}. It consists of an electron drive beam, an electron witness beam, and a singly ionized lithium or argon plasma with a density of $n_0 = 7\times 10^{15}\,\mathrm{cm}^{-3}$. Lithium and argon are considered to illustrate the effect of a nonlinearity owing to ion motion and beam-induced ionization, respectively.
The drive beam is bi-Gaussian with an rms length of $\sigma_{d,z} = 42\,$µm and is located at the origin of the co-propagating coordinate system. We chose an emittance of $\epsilon_{d, [x,y],0} = 60\,$µm~\cite{emittance_explanation}. The witness beam is also bi-Gaussian with initial emittances of $\epsilon_{x,0} = 160\,$µm and $\epsilon_{y,0} = 0.54\,$µm in the horizontal and vertical planes, respectively. Its length is $\sigma_z = 18\,$µm and it is located on axis, $334\,$µm behind the drive beam. The drive and witness beams have initial energies of $31.9\,$GeV ($\gamma_d = 62500$) and $5.1\,$GeV ($\gamma_w = 10000$), charges of $4.3\,$nC and $1.6\,$nC,  and their transverse rms sizes are matched to the blowout wake. The simulations are conducted with the quasi-static, 3D PIC code HiPACE\texttt{++}~\cite{Diederichs:2022:CPC} using its mesh refinement capabilities. The complete numerical settings for all the simulations discussed in this paper are available online~\cite{Diederichs:2024:dataset}. In what follows, $E_0 = m_e c^2 k_p/e$ is the cold, nonrelativistic wavebreaking limit, $k_p = \omega_p/c$ the plasma wavenumber, $\omega_p = \sqrt{n_0 e^2 / (m_e \epsilon_0)}$ the plasma frequency, and $\epsilon_0$ the vacuum permittivity.

\begin{figure}
	\centering
	\includegraphics[trim={0 0 0 0},clip, width=3.375in]{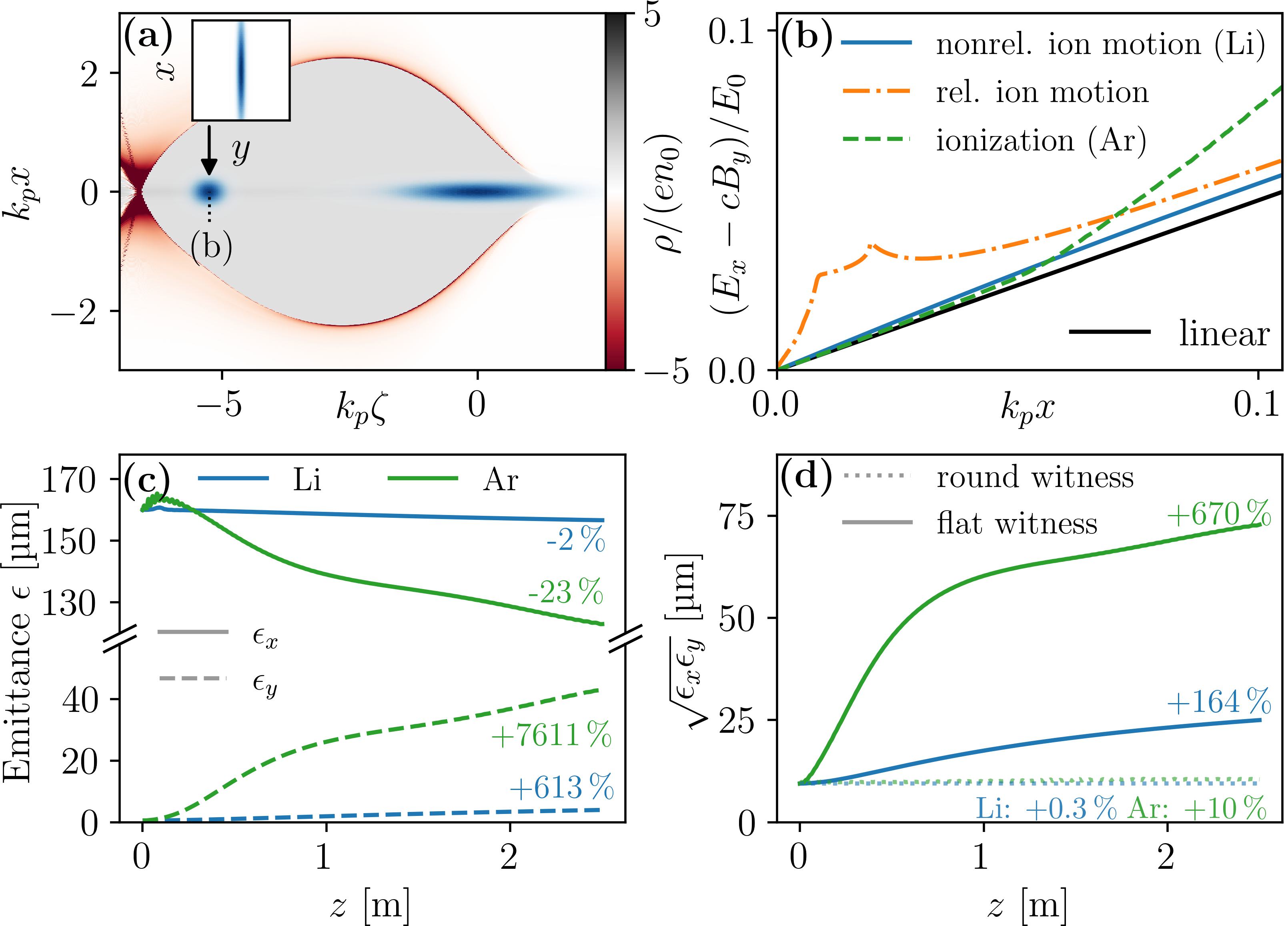}
	\caption{(a) Normalized plasma charge density (grey-red colorscale) and drive and witness beams (blue) in the $x$-$\zeta$ plane, where $\zeta = z - ct$ is the co-moving variable and $c$ the speed of light; inset: transverse profile of the flat beam. (b) Examples of nonlinear transverse wakefields. The blue line (nonrelativistic ion motion) corresponds to a lineout along the dashed line of the case in (a). Other colored lines show wakefields with relativistic ion motion (orange), induced by increasing the initial witness beam energy to $478\,$GeV, thereby decreasing the matched transverse spot size by $\sim 10\times$; drive-beam-induced ionization (dashed green line), obtained with argon. (c) Emittance in $x$ (solid lines) and $y$ planes (dashed lines) and (d) $\sqrt{\epsilon_x\epsilon_y}$ for a flat beam in the nonrelativistic ion motion regime (Li, blue lines) and the ionization regime (Ar, green lines) and for corresponding round beams (dotted lines) with the same initial $\sqrt{\epsilon_x\epsilon_y}$.
    \label{fig:fig1} }
\end{figure}

Figure~\ref{fig:fig1} shows the nonlinear wake and resulting emittance mixing as the witness beam is accelerated from $5.1$ to $\sim 21$\,GeV.
For a flat witness beam, the large horizontal emittance decreases in lithium by $3.4\,$µm, or $-2\,\%$, and in argon by $-23\,\%$. At the same time, the small vertical emittance increases by $3.4\,$µm, or $+613\,\%$ and in argon by $+7611\,\%$. As collider luminosity scales as the inverse of the geometric average of the transverse emittances $\sqrt{\epsilon_x\epsilon_y}$, this quantity is tracked in the rest of this work. In the example of Fig.~\ref{fig:fig1}, $\sqrt{\epsilon_x\epsilon_y}$ increases in lithium by $+164\,\%$ (in argon by $+670\,\%$).
Notably, a round beam matched to the same wakefield with the same initial $\sqrt{\epsilon_x\epsilon_y}$ experiences in lithium a small growth of only $+0.3\,\%$ (in argon by $+10\,\%$) due to mismatch to the nonlinear field, showing that the drive-beam-induced nonlinearity of the transverse field due is the dominant driver of emittance growth in the flat beam case for neither gas. 


In the following, we investigate emittance mixing for the regime of non-relativistic ion motion shown in Fig.~\ref{fig:fig1}, by means of test particles simulations.
We consider a simplified model based on Ref.~\cite{Benedetti:PRAB:2017} that represents well the perturbed transverse wakefield $W_{[x,y]}$ in this regime:
\begin{equation}
\label{eq:coupled_wakes}
W_{[x,y]}= \frac{k_p [x,y] E_0}{2} \left[1 + \alpha_{[x,y]} H\left(\frac{r^2}{2 L_{[x,y]}^2}\right)\right],
\end{equation}
 where $W_x=E_x-cB_y$, and $W_y=E_y+cB_x$ ($E_{[x,y]}$ and $B_{[x,y]}$ are the electric and magnetic fields in the wake, respectively), 
$H(q) = [1-\exp(-q)]/q$, $r = (x^2 + y^2)^{1/2}$ is the radius, and $L_{[x,y]}$ and $\alpha_{[x,y]}$ the characteristic size and amplitude of the nonlinearity. 

\begin{figure}
	\centering
	\includegraphics[trim={0 0 0 0},clip, width=3.375in]{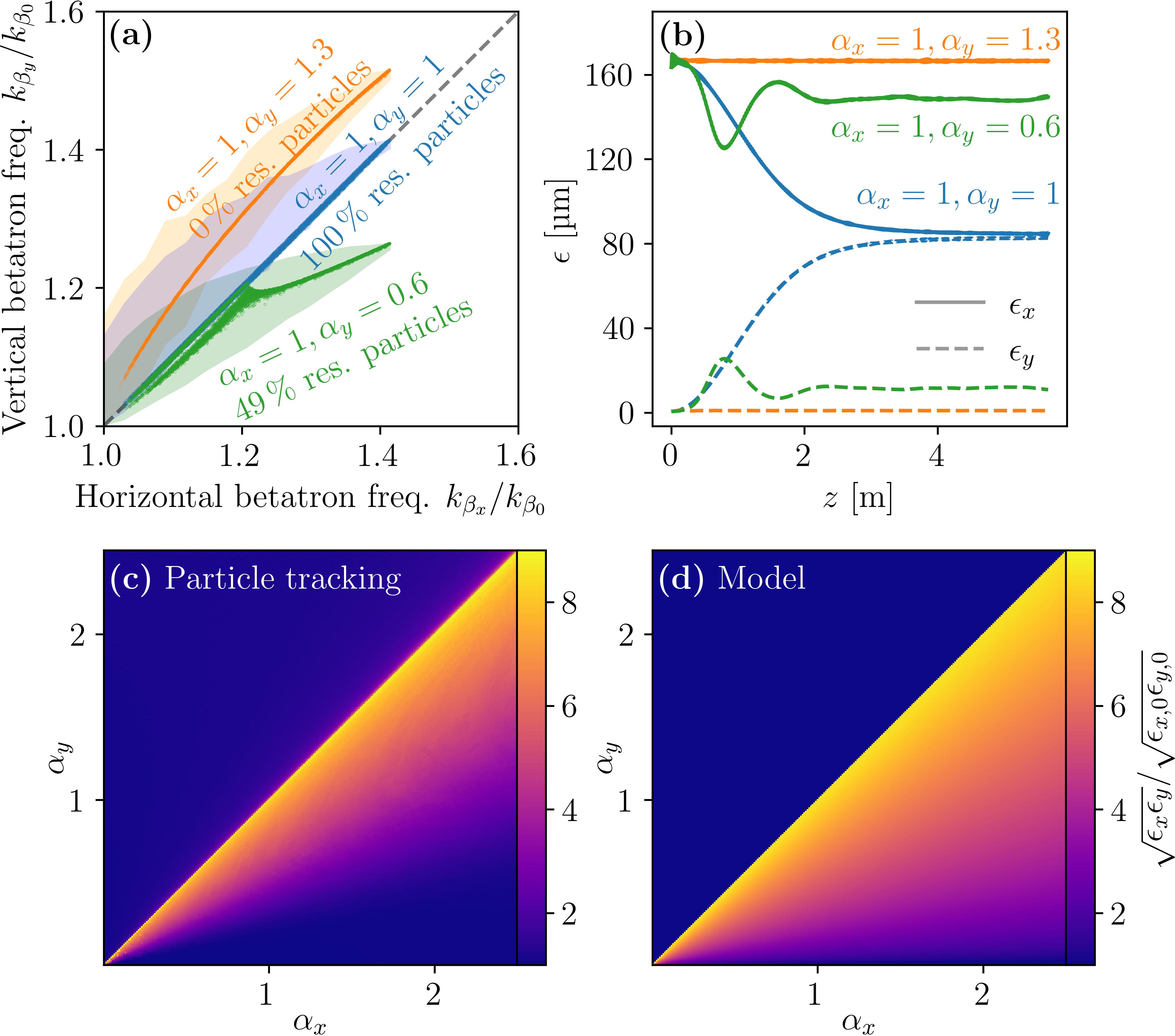}
	\caption{(a) Distribution of betatron frequencies in the $(k_{\beta,x}, k_{\beta,y})$ plane for three different nonlinearity coefficients $\alpha_y = 1$ (blue points), $\alpha_y = 1.3$ (orange points), and $\alpha_y = 0.6$ (green points). For all cases $\alpha_x=1$. The shaded areas denote the initial instantaneous betatron frequencies, the dots show them averaged over many betatron periods. (b) The resulting emittance evolution in $x$ (solid lines) and $y$ (dashed lines). The final $\sqrt{\epsilon_x\epsilon_y}$ at saturation is shown as a function of $\alpha_x$ and $\alpha_y$ for (c) particle tracking and (d) analytical model from Eq.~\ref{eq:emitt_sat_exact}.}
    \label{fig:fig2}
\end{figure}

Beams of test particles with the same properties as the flat witness beam discussed in Fig.~\ref{fig:fig1} are propagated in the transverse wakefields given by Eq.~\eqref{eq:coupled_wakes} (assuming no acceleration) for various nonlinearity coefficients $\alpha_{[x,y]}$ and fixed length scales $L_x = L_y = \sigma_{x,d}$. The length scale was chosen to be the drive beam rms size, since the non-relativistic ion motion is caused by the symmetric drive beam.

The mixing process can be understood by analyzing the single-particle orbits. A beam particle moving in nonlinear wakefields performs transverse betatron oscillations in the $x$ and $y$ planes with frequencies $k_{\beta,x}$ and $k_{\beta,y}$. These frequencies will, in general, differ from the unperturbed betatron frequency $k_{\beta,0}=k_p/(2\gamma_w)^{1/2}$, and depend on the particle's betatron amplitude and, in turn, its initial conditions.  Figure~\ref{fig:fig2}~(a) shows the distribution of betatron frequencies for the beam particles in the three cases considered in Fig.~\ref{fig:fig2}~(b).

In the presence of a coupling term (assumed weak, $\alpha_{[x,y]} H(r^2/2 L_{[x,y]}^2) \lesssim 1$), the $x$ and $y$ orbits form a system of two coupled oscillators.
As generally expected from such a system~\cite{Kovaleva:2013}, particles satisfying the resonance condition $k_{\beta,x} \simeq k_{\beta,y}$, i.e., near the diagonal of Fig.~\ref{fig:fig2}~(a), experience an exchange of their horizontal and vertical betatron oscillation amplitudes. This exchange occurs over a timescale much longer than the betatron period. These resonant particles are responsible for the decrease of $\epsilon_x$ and increase of $\epsilon_y$. In contrast, for particles far from the resonance, the amplitude of the oscillations in $x$ and $y$ are both independently preserved. Overall, the fraction of resonant particles in the beam determines the degree of emittance mixing.
The case $\alpha_{x} = \alpha_y = 1$ (blue lines) has 100\,\% resonant particles, which leads to a full equalization of the emittances. For the case $\alpha_{x} = 1$, $\alpha_y = 1.3$ (orange lines) there are no resonant particles and, hence, no emittance mixing. Finally, in the case of $\alpha_{x} = 1$, $\alpha_y = 0.6$ (green lines), 49\,\% of the particles are resonant, leading to partial mixing. For more details on the last case, see the Supplemental Material.

Given the wakefields in Eq.~\eqref{eq:coupled_wakes}, it is possible to estimate the emittances of a flat beam at saturation $\epsilon_{[x,y]}^*$:
\begin{equation}
\label{eq:emitt_sat_exact}
\begin{cases}
{ \epsilon_{x}^* \simeq \left(1-\frac{\eta_r}2\right)\epsilon_{x,0}}\\

{ \epsilon_{y}^* \simeq \left(1-\eta_r\right)\epsilon_{y,0}+\frac 12 \eta_r \frac {\alpha_y}{\alpha_x}\frac{L_x^2}{L_y^2} \epsilon_{x,0},}
\end{cases}
\end{equation}
where
\begin{equation}
\label{eq:fraction_resonant_particles}
\eta_r=
\begin{cases}
{\exp\left[-\frac{4 k_p^2 L_x^2 L_y^2}{k_{\beta,0}\epsilon_{x,0}} \frac{\alpha_x-\alpha_y}{3\alpha_x L_y^2-2\alpha_y L_x^2}\right] ,\quad \alpha_y \le \alpha_x}\\

{0, \quad \alpha_y>\alpha_x}
\end{cases}
\end{equation}
is the fraction of resonant particles in the beam (see the Supplemental Material for details).

Figure \ref{fig:fig2}~(c) and (d) show the relative growth of the geometric emittance, $(\epsilon_{x}^*\epsilon_{y}^*/\epsilon_{x,0}\epsilon_{y,0})^{1/2}$, in the $(\alpha_x, \alpha_y)$ plane obtained with particle tracking (c), and with the model (d), respectively. 
The model reproduces the main qualitative and quantitative conclusions observations: maximal growth of the geometric emittance is observed for $\alpha_{x} = \alpha_y$ where all beam particles are resonant. When $\alpha_{y} < \alpha_x$, the fraction of resonant particles decreases, resulting in reduced emittance mixing. For $\alpha_{y} > \alpha_x$, no resonant particles are present, and, hence, no emittance growth from mixing is observed. 


Energy transfer between oscillation modes in coupled oscillators at resonance is a general physics process~\cite{Kovaleva:2013}, and can be observed for instance in the Wilberforce pendulum~\cite{Wilberforce:1894, Berg:1991}.
The dynamics in a plasma accelerator considered in this article are complex, but the emittance exchange process itself does not depend on the specific shape of the nonlinear coupling and is therefore observed with any effects causing coupling (e.g., ionization, ion motion, non-uniform plasma density, etc.).
Similar to resonance in RF-based accelerators, determining which particles are trapped in the resonance is a non-trivial task in general~\cite{Guignard:1976,Edwards:2008}, and is best studied with self-consistent numerical simulations.

Emittance mixing due to resonant particles explains the observed drastic emittance increase in Fig.~\ref{fig:fig1}: since the ion motion (lithium) and ionization (argon) are caused by an axisymmetric drive beam, the resulting coupled, nonlinear fields are axisymmetric. Consequently, all witness beam particles share the same betatron frequency in both the $x$ and $y$ directions, making them resonant and leading to the strong emittance mixing observed.

\begin{figure}
	\centering
	\includegraphics[trim={0 0 0 0},clip, width=3.375in]{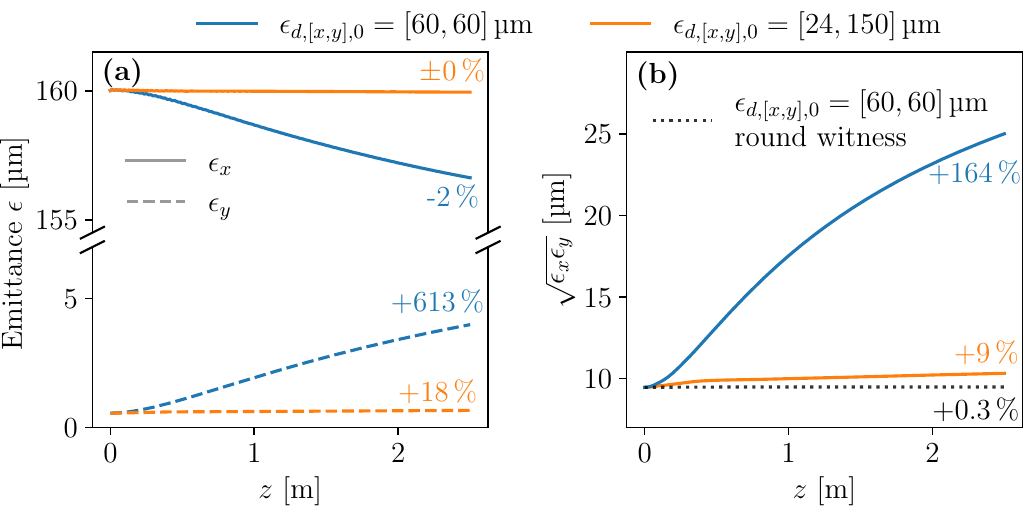}
	\caption{The evolution of emittances (a) $\epsilon_x$ (solid lines) and $\epsilon_y$ (dashed lines), and (b) $\sqrt{\epsilon_x\epsilon_y}$, for a flat witness beam and a round driver (blue line), a flat witness beam and a flat driver (orange line), and a round witness and a round driver (grey dotted line).} \label{fig:fig3} 
\end{figure}

This observation suggests a possible solution to emittance mixing caused by the drive-beam-induced nonlinear fields: using a flat drive beam to induce asymmetric ion motion and detune the resonance. To confirm this, we ran the simulation from Fig.~\ref{fig:fig1} in lithium with a flat driver. The evolution of the witness beam emittances with both round and flat drive beams is shown in Fig.~\ref{fig:fig3}. While the round drive beam with $\epsilon_{d, [x,y],0} = 60\,$µm causes large emittance mixing in the witness beam, a flat drive beam with emittances $\epsilon_{d, [x,y],0} = [24\,$µm$, 150\,$µm] causes asymmetric ion motion and suppresses the emittance mixing.
The flat driver causes significant ion motion in the $y$ direction and negligible one in $x$. The emittance growth in the $y$ direction for the witness beam can be attributed to mismatch in these nonlinear fields rather than emittance mixing, because the emittance does not decrease in the $x$ direction. This growth could therefore be prevented by nonlinearly matching the witness beam in $y$ to the nonlinear fields~\cite{Benedetti:PRAB:2017, Benedetti:PoP:2021}. Using a flat drive beam is a viable way to suppress emittance mixing due to drive-beam-induced ion motion, since it breaks the symmetry and, in this case, the resonance.

When the nonlinearity is created by the driver, as above, emittance exchange in the witness beam can be mitigated by shaping the driver. Independently, and regardless of the driver (such that this also applies to laser-driven plasma accelerators), a high charge, low emittance, and high energy witness beam can itself trigger nonlinearities resulting in emittance growth. This is the case in particular in the regime relevant for a multi-TeV collider, where the flat witness beam creates nonlinear coupled wakefields due to ion motion or ionization. The induced nonlinearities are not symmetric because the flat witness beam is not symmetric. Nevertheless, resonance still occurs for enough particles to cause considerable emittance mixing.
To assess the emittance mixing in this regime, we choose the same parameters as in Fig.~\ref{fig:fig1} and reduce the witness beam emittance to a level relevant for a future multi-TeV-class collider, namely $\epsilon_{[x,y],0} = [5\,$µm$, 35\,\mathrm{nm}]$. 
At a plasma density of $n_0 = 7\times 10^{15}\,\mathrm{cm}^{-3}$, the space charge fields of this matched beam at $\gamma_w = 10000$ exceed $300\,$GV/m, leading to relativistic ion motion. To prohibit ionization effects of the background ions, fully ionized hydrogen is used, although another sufficiently ionized element could be considered. The drive beam is assumed rigid with a large enough emittance not to cause ion motion or ionization to isolate the effects of the witness-beam-induced ion motion.
The witness beam is propagated over a distance of $77.5\,$m and gains energy from $5.1$ to $500$\,GeV.

\begin{figure}
	\centering
	\includegraphics[trim={0 0 0 0},clip, width=3.375in]{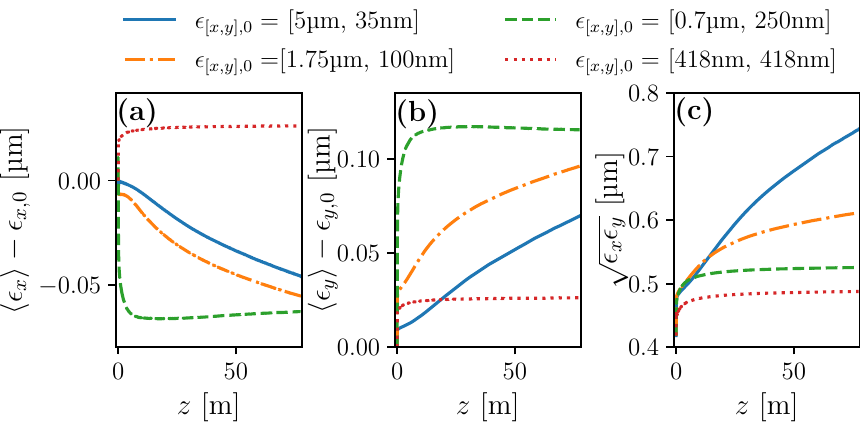} 
	\caption{Evolution of emittances (a) $\epsilon_x$, (b) $\epsilon_y$, and (c) $\sqrt{\epsilon_x\epsilon_y}$, with different initial values for the horizontal and vertical emittances for the witness bunch, but the same initial $\sqrt{\epsilon_x\epsilon_y}$. The average slice emittance (subtracted by the initial emittance) is used in (a) and (b) to avoid head-to-tail mismatches that lead to an increase in the emittance in $x$ and mask the mixing, while (c) shows $\sqrt{\epsilon_x\epsilon_y}$.}
    \label{fig:fig4}
\end{figure}

Figure \ref{fig:fig4} shows the emittance evolution from four different witness beams with decreasing aspect ratio (from very flat to round with the same initial $\sqrt{\epsilon_x\epsilon_y}$): $\epsilon_{x,0} = 5$\,µm, 1.75\,µm, 0.7\,µm, and $418\,\mathrm{nm}$, while $\epsilon_{y,0} = 35\,\mathrm{nm}$, $100\,\mathrm{nm}$, $250\,\mathrm{nm}$, and $418\,\mathrm{nm}$, respectively.
In all flat beam cases, the emittance in $x$ decreases while it increases in $y$. A flatter beam results in a larger growth of $\sqrt{\epsilon_x\epsilon_y}$, and the emittances of the flattest beams do not reach saturation after acceleration to 500\,GeV. The emittance of the round beam increases in both planes by $6\,\%$ due to the nonlinearity of the ion-motion-perturbed fields, which, again, could be avoided by perfectly matching the beam to the nonlinear fields~\cite{Benedetti:PRAB:2017, Benedetti:PoP:2021}. 
The emittances of the flat beams increase by $26\,\%$, $46\,\%$, and $78\,\%$, respectively.

In this Letter we have shown that nonlinear transverse wakefields in plasma accelerators couple the particle dynamics in the transverse planes and cause emittance mixing. If not mitigated, this effect can make a flat beam round resulting in drastic reduction in luminosity. The mixing is due to a resonance between the betatron oscillations in the horizontal and vertical planes, and the most violent emittance mixing occurs for an axisymmetric nonlinear transverse wakefield. This effect can be mitigated by breaking the cylindrical symmetry, which reduces the fraction of resonant particles and hence the emittance exchange. For nonlinearities caused by the driver, this can be achieved by tailoring the driver properties with, e.g., flat drive beams or laser drivers (for which ion motion is negligible). Nonlinearities caused by a strong witness beam (high charge, high energy, low emittance), as found in current plasma-based collider designs, also result in considerable beam degradation. This problem needs to be addressed for future specific designs relying on flat witness beams. 

The choice of element used to generate the plasma is decisive: emittance mixing can be controlled by avoiding both ion motion (stronger for elements with larger charge-to-mass ratio, i.e., light elements or heavy elements ionized to high levels) and beam-induced ionization (stronger for heavy elements when not sufficiently pre-ionized). While only the blowout regime was discussed in detail, emittance mixing can also occur in other regimes such as the linear and quasi-linear regimes~\cite{Esarey:2009} as well as plasma-based positron acceleration schemes~\cite{Zhou:2021, Lotov:2007, Silva:2021, Diederichs:2019} that operate with coupled, nonlinear focusing fields. Although hollow core plasma channels~\cite{Schroeder:2023} could, in principle, be used to avoid emittance mixing, flat beams are then susceptible to beam breakup due to a self-induced quadrupole moment~\cite{Zhou:2021}.

\begin{acknowledgments}
We acknowledge fruitful discussions with Reinhard Brinkmann, Carl A. Lindstr\o m, Ming Zeng, Rob Shalloo and Eric Esarey.
We acknowledge the Funding by the Helmholtz Matter and Technologies Accelerator Research and Development Program.
This work was supported by the Director, Office of Science, Office of High Energy Physics, of the U.S. Department of Energy, under Contract No. DE-AC02-05CH11231, and used the computational facilities at the National Energy Research Scientific Computing Center (NERSC).
We gratefully acknowledge the Gauss Centre for Supercomputing e.V. (www.gauss-centre.eu) for funding this project by providing computing time through the John von Neumann Institute for Computing (NIC) on the GCS Supercomputer JUWELS at J\"ulich Supercomputing Centre (JSC). This research was supported in part through the Maxwell computational resources operated at Deutsches Elektronen-Synchrotron DESY, Hamburg, Germany.
This work was funded by the Deutsche Forschungsgemeinschaft (DFG, German Research Foundation)—491245950 
\end{acknowledgments}

\bibliography{Bibliography}

\pagebreak
\clearpage

\begin{center}
\textbf{\large Supplemental Material: Resonant emittance mixing of flat beams in plasma accelerators}
\end{center}

\setcounter{equation}{0}
\setcounter{figure}{0}
\setcounter{table}{0}
\setcounter{page}{1}
\makeatletter
\renewcommand{\theequation}{S\arabic{equation}}
\renewcommand{\thefigure}{S\arabic{figure}}

\section{Trapping in the resonance}
\label{sec:trapping}
\begin{figure}
	\centering
	\includegraphics[trim={0 0 0 0},clip, width=3.375in]{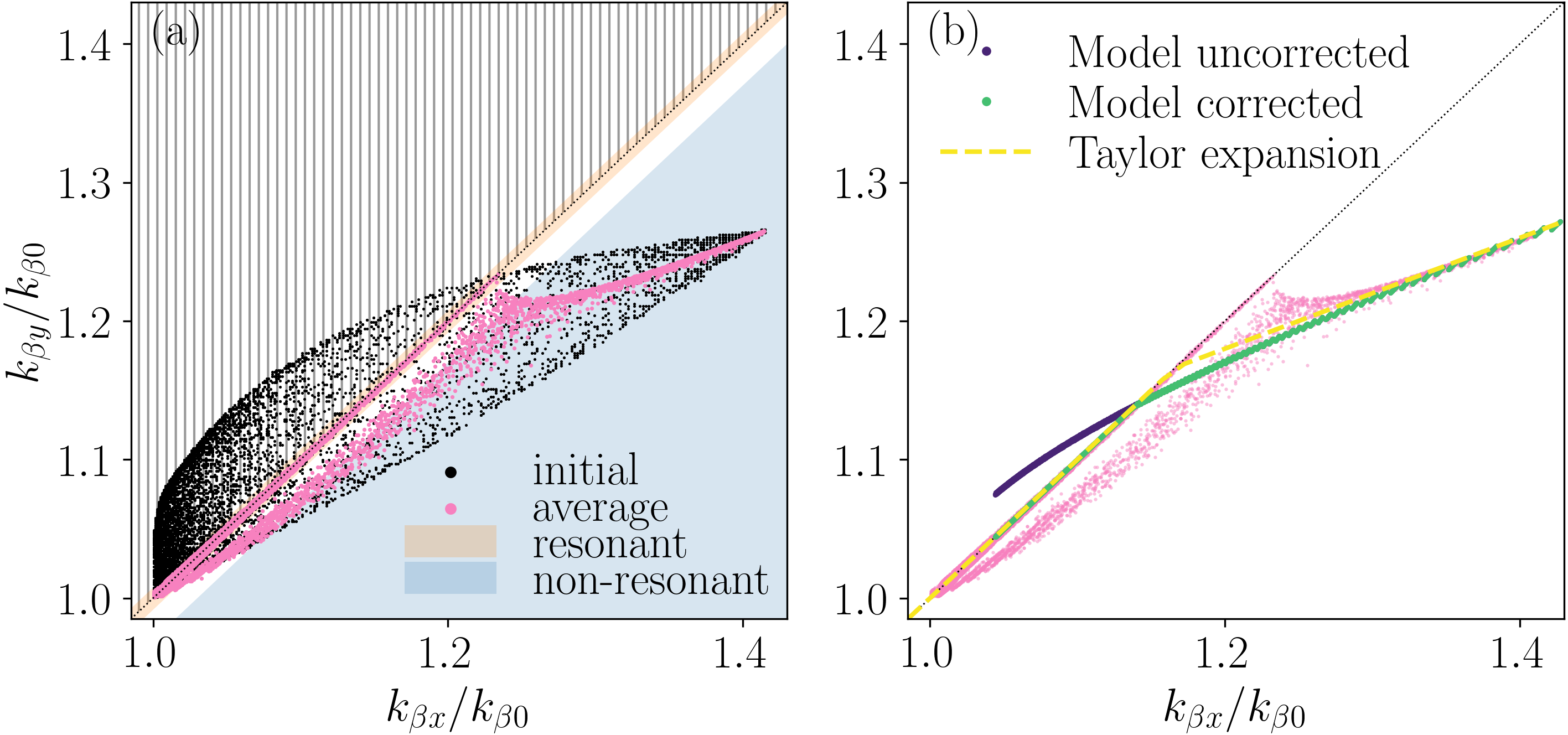}
	\caption{(a) Initial instantaneous betatron frequencies distribution (black), and betatron frequencies distribution averaged over a large number of betatron periods (pink). Particles whose average betatron frequencies are near the diagonal (orange area) are resonant. For these external fields ($\alpha_x = 1.0$ and $\alpha_y=0.6$), all the particles with initial instantaneous betatron frequency above the diagonal, denoted by the hatches, are trapped in the resonance. Particles far enough below the diagonal (blue area) are non-resonant. (b) Initial betatron frequencies from the model before (black) and after (green) correction. The correction consists in projecting all particles above the diagonal onto the diagonal. The Taylor expansion is shown as a dashed yellow line, where the same correction has been applied. The average tune map from (a) is reproduced here for easier comparison.}
    \label{fig:figa1}
\end{figure}

Figure~\ref{fig:figa1}~(a) shows the distribution of betatron frequencies (tune map) for the green case of Fig.~2~(a): $\alpha_x = 1.0$ and $\alpha_y=0.6$. The betatron frequency in x (resp. y) is measured for each test particle as the inverse of the distance between two consecutive maxima in the x (resp. y) orbits.

Because of the coupling between the transverse directions, the amplitude of betatron oscillations in $x$ and in $y$ is not constant, and evolves over a distance much longer than the betatron period. Consequently, the betatron frequencies in $x$ and $y$ also evolve slowly during the propagation. The initial betatron frequency (black dots) exhibit a relatively smooth distribution, consistent with the Gaussian initial distributions in position and momentum. When averaged over a long propagation distance (pink dots), structures appear in the frequency map as a significant fraction of particles gathers near the diagonal. These are the particles with initial instantaneous betatron frequency satisfying $k_{\beta y} \geq k_{\beta x}$ (i.e., those for which the black dots are above the diagonal, as denoted by the hatches). Resonant particles are those with average betatron frequencies near the diagonal (orange area), non-resonant particles have average betatron frequencies far enough below the diagonal (blue area).

For the sake of simplicity, in Fig.~2 of the article, all particles \emph{not} in the blue area are counted as resonant. Note that the location of the hatched region depends on the external fields: for $\alpha_x = 1.3$ and $\alpha_y=1.0$, it would consist of the lower right corner $k_{\beta y} \leq k_{\beta x}$ , while for $\alpha_x = 1.0$ and $\alpha_y=1.0$ it would consist of the whole plane. Figure~\ref{fig:figa1}~(b) shows the betatron frequencies from the model (see Sec.~\ref{sec:model}), demonstrating good agreement with simulations. The trapping in the resonance is imposed as a phenomenological correction to the model.

\section{Analytical model for emittance mixing in flat beams}
\label{sec:model}
Emittance mixing in a flat witness beam can be understood in terms of the resonance of betatron oscillations in the horizontal and vertical planes when a nonlinearity in the wakefield couples the motion in these planes.

We consider the following transverse equations of motion for the witness beam particles (written in a compact form),
\begin{equation}
\label{eq:eq_motion}
\begin{cases}
\displaystyle \frac{d [x,y]}{dz}=\frac{u_{[x,y]}}{\gamma_w} \\
{}\\
{\displaystyle \frac{d u_{[x,y]}}{dz}}=\displaystyle -\frac{ k_p^2 [x,y]}{2} \left[1+\alpha_{[x,y]} H\left( \frac{r^2}{2L_{[x,y]}^2}\right)\right],
\end{cases}
\end{equation}
where $z$ is the propagation distance ($z=ct$, with $t$ the time), $x$ [$y$] is the transverse coordinates of the particle in the horizontal [vertical] plane, $u_x$ [$u_y$] the horizontal [vertical] component of the transverse momentum normalized to $m_e c$, $r^2=x^2+y^2$, $H(q)=[1-\exp(-q)]/q$, $\gamma_w\gg 1$ the relativistic factor for the beam, 
$\alpha_{[x,y]}$ the nonlinearity coefficients, and $L_{[x,y]}$ the characteristic sizes of the nonlinearity. 

We introduce action-angle coordinates, defined as
\begin{equation}
\label{eq:aa_variable}
\begin{cases}
{\displaystyle [x,y]=\left(\frac{2j_{[x,y]}}{\gamma_w k_{\beta,0}}\right)^{1/2}\cos \theta_{[x,y]}} \\ 
{}\\
{\displaystyle u_{[x,y]}=-\left(2\gamma_w k_{\beta,0} j_{[x,y]} \right)^{1/2}\sin \theta_{[x,y]}},  
\end{cases}
\end{equation}
where $j_x$ [$j_y$] and $\theta_x$ [$\theta_y$] are, respectively, the action and angle coordinates in the horizontal [vertical] plane, and $k_{\beta,0}=k_p/(2\gamma_w)^{1/2}$ is the linear (i.e., unperturbed) betatron wavenumber. Note that the horizontal and vertical action variables for a particle are related to the amplitude of the betatron oscillations in the corresponding plane. Motion of beam particles is easier to analyze using these coordinates, as opposed to Cartesian coordinates, since the symmetries of the problem (e.g., quasi-periodic motion) become explicit.

Rewriting the equations of motion Eq.~\eqref{eq:eq_motion} using action-angle coordinates we obtain
\begin{equation}
\label{eq:action-angle} 
\begin{cases}
{\displaystyle\frac{d\theta_{[x,y]}}{dz}=  \displaystyle k_{\beta,0}\left[1+\alpha_{[x,y]} \cos^2 \theta_{[x,y]} H\left( \frac{r^2}{2L_{[x,y]}^2}\right)\right]}  \\
{}\\
{\displaystyle\frac{dj_{[x,y]}}{dz}=\displaystyle \alpha_{[x,y]} k_{\beta,0}j_{[x,y]} \sin 2\theta_{[x,y]} H\left( \frac{r^2}{2L_{[x,y]}^2}\right),} 
\end{cases}
\end{equation}
where $r^2=2(j_x \cos^2\theta_x+j_y\cos^2\theta_y)/\gamma_w k_{\beta,0}$. We see that the system can be described as a set of nonlinearly-coupled oscillators. 

For the initial condition considered in this work (i.e., a Gaussian beam distribution linearly matched in the unperturbed wakefield with $\epsilon_{[x,y],0}$ the rms emittances on each plane) the initial angles, $\theta_{[x,y],0}$, are uniformly distributed in $[0, 2\pi]$, and the initial actions, $j_{[x,y],0}$, are distributed as 
\begin{equation}
\label{eq:j_distrib}
f(j_{x,0}, j_{y,0}) =\frac{1}{\epsilon_{x,0}\epsilon_{y,0}} \exp\left(-\frac{j_{x,0}}{\epsilon_{x,0}}-\frac{j_{y,0}}{\epsilon_{y,0}}\right).
\end{equation}

The nonlinearity in the wakefield results in amplitude-dependent (i.e., action-dependent) betatron frequencies for the beam particles. Following canonical perturbation theory, an estimate for the initial (i.e., valid before significant evolution occurs) frequencies can be obtained by averaging over the angular coordinates the expressions for $d\theta_{[x,y]}/dz$ in Eq.~\eqref{eq:action-angle}, namely 
\begin{equation}
\label{eq:initial_tune_def}
k_{\beta,[x,y],0}=\frac 1 {(2\pi)^2} \int \int \left(\frac{d \theta_{[x,y]}}{dz}\right) d\theta_x d\theta_y.
\end{equation}
For a flat beam, $\epsilon_{y,0}\ll \epsilon_{x,0}$, we have that, in general, $j_{y,0}\ll j_{x,0}$, and so the betatron frequencies are determined solely by the initial value of the particle's action in the horizontal plane. We obtain,
\begin{equation}
\label{eq:betatron_exact}
\begin{cases}
{\displaystyle \frac{k_{\beta,x,0}}{k_{\beta,0}}\simeq 1+\alpha_x Q_2 (j_{x,0}; L_x)}\\ 
{}\\
{\displaystyle \frac{k_{\beta,y,0}}{k_{\beta,0}}\simeq 1+\frac{\alpha_y}2 Q_0 (j_{x,0}; L_y),}
\end{cases}
\end{equation}
where 
\begin{equation}
Q_p (j_{x,0}; L)=\frac 1{2\pi} \int_0^{2\pi}  (\cos\varphi)^p H\left[\frac{2k_{\beta,0}j_{x,0}\cos\varphi}{(k_pL)^2}\right]d \varphi.
\end{equation}
Note that the expressions in Eq.~\eqref{eq:betatron_exact} are valid for $\alpha_{[x,y]}\lesssim 1$. 
From Eq.~\eqref{eq:betatron_exact} we see that both the horizontal and vertical betatron frequencies decrease as the value of the horizontal action increases.
The black line in Fig.~\ref{fig:figa1}(b) shows the initial distribution of betatron frequencies computed using Eq.~\eqref{eq:betatron_exact} for $\alpha_x=1$, $\alpha_y=0.6$ (the other beam and plasma parameters are $n_0=7\times 10^{15}$ cm$^{-3}$, $L_x=L_y=6$ $\mu$m, $\gamma_w=10000$, $\epsilon_{x,0}=160$ $\mu$m, and $\epsilon_{y,0}=0.54$ $\mu$m). 

If the characteristic size of the nonlinearity is (much) larger than the characteristic size of the beam in the horizontal plane, i.e., $\sigma_{x}\sim ( \epsilon_{x,0} / \gamma_w k_{\beta,0})^{1/2}\ll L_{[x,y]}$ (we refer to this as the small beam limit), then $k_{\beta,0}j_{x,0}\ll (k_pL_{[x,y]})^2$, and so we can use the approximation $H(q)\simeq 1-q/2$. In this limit the expressions for the betatron frequencies Eq.~\eqref{eq:betatron_exact} simplify to
\begin{equation}
\label{eq:betatron_taylor}
\begin{cases}
{\displaystyle \frac{k_{\beta,x,0}}{k_{\beta,0}}\simeq 1+\frac{\alpha_x}2\left[1-\frac 3 4 \frac{k_{\beta,0} j_{x,0}}{(k_pL_x)^2}\right]}\\
{}\\
{\displaystyle \frac{k_{\beta,y,0}}{k_{\beta,0}}\simeq 1+\frac{\alpha_y}2\left[1-\frac 1 2 \frac{k_{\beta,0} j_{x,0}}{(k_pL_y)^2}\right].}
\end{cases}
\end{equation}

Resonant particles are the ones for which the horizontal and vertical betatron frequencies are equal. (More generally, a resonance is present if the ratio between the horizontal and vertical betatron frequencies is a rational number. Here we consider the simplest case where the ratio between the frequencies is 1.) 
The value of the horizontal action for which a beam particle is (initially) resonant, $j_{x,0}^{(r)}$, can be obtained by solving the equation 
\begin{equation}
k_{\beta,x,0}(j_{x,0}^{(r)}) = k_{\beta,y,0}(j_{x,0}^{(r)}).
\end{equation}
In the small beam limit, and for $L_x \approx L_y$, a finite, non-negative solution to this equation exists for $\alpha_y \le \alpha_x$, and reads
\begin{equation}
\label{eq:sb_jr}
j_{x,0}^{(r)}\simeq \frac{4 k_p^2 L_x^2 L_y^2}{k_{\beta,0}}  \frac{\alpha_x-\alpha_y}{3\alpha_x L_y^2-2\alpha_y L_x^2}.\end{equation}
Note that the positive solution for $\alpha_y>(3/2) (L_y/L_x)^2\alpha_x$ is not acceptable since it violates the small beam approximation used to derive Eq.~\eqref{eq:sb_jr}. 

For values of the initial horizontal action larger than $j_{x,0}^{(r)}$, Eq.~\eqref{eq:betatron_exact} predicts that the betatron frequency will cross the resonance line. However, a direct numerical solution to the equations of motion [Eq.~\eqref{eq:eq_motion} or Eq.~\eqref{eq:action-angle}] indicates, instead, that these particles are trapped in the resonance, with an effective vertical betatron frequency that is lower than the theoretical one, and equal to the horizontal one (see the previous Sec.~\ref{sec:trapping}). 
Based on these findings, and starting from the knowledge of the initial betatron frequencies, we can build a 
model for the distribution of betatron frequencies of the beam particles which is valid at later times (asymptotic), namely
\begin{equation}
\label{eq:corrected_tunes}
\begin{cases}
 {k_{\beta, x}=k_{\beta,x,0}}\\
 {}\\
 {k_{\beta,y}
=\begin{cases}
{k_{\beta,y,0}\quad  \hbox{for} \quad j_{x,0}<j_{x,0}^{(r)}}\\
{k_{\beta,x,0}\quad  \hbox{for} \quad j_{x,0} \geq j_{x,0}^{(r)}}.
\end{cases} 
 }
\end{cases}    
\end{equation}
The green line in Fig.~\ref{fig:figa1}(b) shows the  asymptotic distribution of betatron frequencies given by Eq.~\eqref{eq:corrected_tunes}. The yellow dashed line is the asymptotic distribution in the small beam limit (Taylor expansion). We see that Eq.~\eqref{eq:corrected_tunes} provides a good description of the beam's asymptotic betatron footprint.

The fraction of beam particles in the resonance can be computed considering that, according to our model, particles for which $j_{x,0} \ge j_{x,0}^{(r)}$ are resonant, and taking into account that the initial distribution of actions is given by Eq.~\eqref{eq:j_distrib}. Namely, 
\begin{eqnarray}
\label{eq:fraction_of_resonant_particles}
\eta_r &=& \int_{j_{x,0}^{(r)}}^{\infty} dj_{x,0}\int_0^{\infty} dj_{y,0}\, f(j_{x,0}, j_{y,0})  \nonumber\\
&=& \exp\left(-\frac{j_{x,0}^{(r)}}{\epsilon_{x,0}}\right).
\end{eqnarray}
For the cases shown in Fig.~2~(a) in the Letter, Eq.~\eqref{eq:fraction_of_resonant_particles} predicts $\eta_r=0.64$ for $\alpha_x=1, \alpha_y=0.6$, $\eta_r=1$ for $\alpha_x=\alpha_y=1$, and $\eta_r=0$ for $\alpha_x=1, \alpha_y=1.3$. These values are in good qualitative  agreement with those obtained in the test particle simulations, which are 0.49, 1.0, and 0.0, respectively.
We can also evaluate the initial average value of the actions in the horizontal and vertical planes for resonant and non-resonant particles. These are given, respectively, by
\begin{eqnarray}
\label{eq:average_jx0_resonant}
\langle j_{x,0}\rangle_{\hbox{\tiny res.}} &=&
\int_{j_{x,0}^{(r)}}^{\infty} dj_{x,0} \int_0^{\infty} dj_{y,0}\, j_{x,0}\, f(j_{x,0}, j_{y,0}) \nonumber\\
&= & \eta_r (\epsilon_{x,0}+j_{x,0}^{(r)}),
\end{eqnarray}
\begin{eqnarray}
\label{eq:average_jy0_resonant}
\langle j_{y,0}\rangle_{\hbox{\tiny res.}} &=& 
\int_{j_{x,0}^{(r)}}^{\infty}  dj_{x,0} \int_0^{\infty} dj_{y,0}\, j_{y,0}\, f(j_{x,0}, j_{y,0}) \nonumber\\
&=&\eta_r\epsilon_{y,0} ,
\end{eqnarray}
\begin{eqnarray}
\label{eq:average_jx0_nonresonant}
\langle j_{x,0}\rangle_{\hbox{\tiny non-res.}} &=&
\int_0^{j_{x,0}^{(r)}} dj_{x,0} \int_0^{\infty} dj_{y,0}\, j_{x,0}\, f(j_{x,0}, j_{y,0})\nonumber\\
&=& (1-\eta_r)\epsilon_{x,0} - \eta_r j_{x,0}^{(r)}.
\end{eqnarray}
and
\begin{eqnarray}
\label{eq:average_jy0_nonresonant}
\langle j_{y,0}\rangle_{\hbox{\tiny non-res.}} &=& 
\int_0^{j_{x,0}^{(r)}} dj_{x,0} \int_0^{\infty} dj_{y,0}\, j_{y,0}\, f(j_{x,0}, j_{y,0})
\nonumber\\
&=&(1-\eta_r)\epsilon_{y,0} .
\end{eqnarray}


The temporal evolution of the angle coordinates for each beam particle can be obtained solving the first equation in Eq.~\eqref{eq:action-angle}. Neglecting oscillations on the $\sim k_{\beta,0}^{-1}$ scale, we obtain
\begin{equation}
\label{eq:angle_solution}
\theta_{[x,y]}(z)\simeq \theta_{[x,y],0}+k_{\beta, [x,y]} z,
\end{equation}
where $k_{\beta,[x,y]}$ are given by Eq.~\eqref{eq:corrected_tunes}.

The temporal evolution of the action coordinates is different for resonant and non-resonant particles. This can be seen rewriting the evolution equation for the action (second equation in Eq.~\eqref{eq:action-angle}) in the small beam limit, by using Eq.~\eqref{eq:angle_solution} for the evolution of the angles, and by averaging over oscillations at the $\sim k_{\beta,0}^{-1}$ scale.
For non-resonant particles, i.e., $k_{\beta,x} \neq k_{\beta,y}$, we have 
\begin{equation}
\frac{dj_{[x,y]}}{dz}\simeq 0,
\end{equation}
and so the actions in both planes are separately \hbox{(quasi-)preserved} during evolution, namely
\begin{equation}
\label{eq:j_nonresonant}
j_{[x,y]}(z) \simeq j_{[x,y],0}.
\end{equation}
For resonant particles, i.e., $k_{\beta,x}= k_{\beta,y}$, we have
\begin{equation}
\label{eq:action_evol_simpl}
\begin{cases}
{\displaystyle \frac{dj_x}{dz}\simeq -k_{\beta,0}^2\frac{\alpha_x}4 \frac{j_x j_y}{(k_p L_x)^2}\sin 2(\theta_{x,0}-\theta_{y,0}) }\\
{}\\
{\displaystyle \frac{dj_y}{dz}\simeq k_{\beta,0}^2\frac{\alpha_y}4 \frac{j_x j_y}{(k_p L_y)^2}\sin 2(\theta_{x,0}-\theta_{y,0}), }
\end{cases}
\end{equation}
and so the actions in the two planes are not separately conserved since the nonlinearity provides a mechanism to couple the particle motion in the horizontal and vertical planes. However, since $(\alpha_y/L_y^2) (d j_x/dz)+(\alpha_x/L_x^2) (d j_y/dz)\simeq 0$, we have that the quantity
\begin{equation}
\label{eq:first_integral}
\frac{\alpha_y}{L_y^2} j_x(z) + \frac{\alpha_x}{L_x^2} j_y(z) = \frac{\alpha_y}{L_y^2} j_{x,0} + \frac{\alpha_x}{L_x^2} j_{y,0} = \hbox{const.}, 
\end{equation}
is (quasi-)preserved, and the value of the constant is set by the initial condition for the actions. Using Eq.~\eqref{eq:first_integral} in Eq.~\eqref{eq:action_evol_simpl} we obtain the following equation describing the evolution of $j_x$, 
\begin{eqnarray}
\label{eq:jx_alone}
\displaystyle \frac{dj_x}{dz} &\simeq& -\frac{k_{\beta,0}^2 \sin 2(\theta_{x,0}-\theta_{y,0})}{4 } \\
\displaystyle&\times&j_x \left[ \frac{\alpha_x}{(k_pL_x)^2} j_{y,0}+\frac{\alpha_y}{(k_pL_y)^2} (j_{x,0}-j_x)\right] . \nonumber
\end{eqnarray}
The asymptotic solution of this equation can be obtained as follows. If $\sin 2(\theta_{x,0}-\theta_{y,0})>0$ (case I), then $d j_x/dz<0$ and so $j_x$ decreases. The horizontal action will continue to decrease (the term within the square parenthesis is always positive) and will tend to the asymptotic value 
\begin{equation}
\label{eq:res_1}
j_x^{*,+}=0, 
\end{equation}
which is a fixed point for Eq.~\eqref{eq:jx_alone}. At the same time Eq.~\eqref{eq:first_integral} implies that $j_y$ will increase and tend to the asymptotic value 
\begin{equation}
\label{eq:res_2}
j_y^{*,+} = j_{y,0}+\frac{\alpha_y}{\alpha_x} \frac{L_x^2}{L_y^2}j_{x,0} \simeq \frac{\alpha_y}{\alpha_x}\frac{L_x^2}{L_y^2} j_{x,0}, 
\end{equation}
where the last equality is valid for a flat beam. Conversely, if $\sin 2(\theta_{x,0}-\theta_{y,0})<0$ (case II), the opposite behaviour is observed: the horizontal emittance will tend to the asymptotic value 
\begin{equation}
\label{eq:res_3}
j_x^{*,-}=j_{x,0}+\frac{\alpha_x}{\alpha_y} \frac{L_y^2}{L_x^2}  j_{y,0}\simeq j_{x,0},
\end{equation}
while the vertical emittance approaches the asymptotic value 
\begin{equation}
\label{eq:res_4}
j_y^{*,-}=0. 
\end{equation}
Note that since $\theta_{x,0}-\theta_{y,0}$ is uniformly distributed in $[0, 2\pi]$, the solutions corresponding to case I and case II occur with the same probability. 


We can now compute the beam emittance at saturation when mixing occurs. We recall that the beam emittance in each plane is given by the average value of the action in that plane if the angle coordinates are uniformly distributed. By separating the contribution of resonant and non resonant particles we have 
\begin{equation}
\label{eq:sat_emitt}
\epsilon_{[x,y]}^* \simeq \langle j^*_{[x,y]}\rangle_{\hbox{\tiny non-res.}}+ \langle j^*_{[x,y]}\rangle_{\hbox{\tiny res.}},
\end{equation}
where $\langle j^*_{[x,y]}\rangle_{\hbox{\tiny non-res.}}$ and $\langle j^*_{[x,y]}\rangle_{\hbox{\tiny res.}}$ are the average values of the action at saturation for non-resonant and resonant particles, respectively. 
For non-resonant particles, using Eq.~\eqref{eq:j_nonresonant} (i.e., $j^*_{[x,y]}=j_{[x,y]}(z) \simeq j_{[x,y],0}$), 
we have
\begin{equation}
\label{eq:j_sat_nonres}
\begin{cases}
{\langle j^*_{x}\rangle_{\hbox{\tiny non-res.}}\simeq \langle j_{x,0}\rangle_{\hbox{\tiny non-res.}} }\\
{}\\
{\langle j^*_{y}\rangle_{\hbox{\tiny non-res.}}\simeq \langle j_{y,0}\rangle_{\hbox{\tiny non-res.}} }.
\end{cases}
\end{equation}
For resonant particles, using Eqs.~\eqref{eq:res_1}, \eqref{eq:res_2}, \eqref{eq:res_3}, \eqref{eq:res_4}, 
we have
\begin{equation}    
\label{eq:j_sat_res}
\begin{cases}
{\displaystyle \langle j^*_x\rangle_{\hbox{\tiny res.}} \simeq\frac 12 \langle j_{x,0} \rangle_{\hbox{\tiny res.}}} \\
{}\\
{\displaystyle\langle j^*_y\rangle_{\hbox{\tiny res.}}\simeq\frac 12 \frac {\alpha_y}{\alpha_x} \frac{L_x^2}{L_y^2} \langle j_{x,0} \rangle_{\hbox{\tiny res.}}. }
\end{cases}
\end{equation}
By inserting Eqs.~\eqref{eq:j_sat_nonres} and \eqref{eq:j_sat_res} into Eq.~\eqref{eq:sat_emitt}, we obtain the following expression for the emittances at saturation
\begin{equation}
\label{eq:emitt_sat_exact}
\begin{cases}
{\displaystyle \epsilon_{x,0}^* \simeq \left(1-\frac{\eta_r}2\right)\epsilon_{x,0}-\frac 12 \eta_r j_{x,0}^{(r)}}\\
{}\\
{\displaystyle \epsilon_y^* \simeq (1-\eta_r)\epsilon_{y,0}+\frac 12 \eta_r \frac {\alpha_y}{\alpha_x}\frac{L_x^2}{L_y^2} (\epsilon_{x,0}+j_{x,0}^{(r)}).}
\end{cases}
\end{equation}
We see that, whenever resonant particles are present (i.e, $j_{x,0}^{(r)}$ is non negative and so $\eta_r >0$), mixing occurs: the horizontal beam emittance decreases, while the vertical one increases.
Equation~\eqref{eq:emitt_sat_exact} can be further simplified considering that, for the parameters considered in this work, $\epsilon_{x,0}$ is larger than $j_{x,0}^{(r)}$, and so we obtain 
\begin{equation}
\label{eq:emitt_sat_final}
\begin{cases}
{\displaystyle \epsilon_x^* \simeq \left(1-\frac{\eta_r}2\right)\epsilon_{x,0}}\\
{}\\
{\displaystyle \epsilon_y^* \simeq (1-\eta_r)\epsilon_{y,0}+\frac 12 \eta_r \frac {\alpha_y}{\alpha_x}\frac{L_x^2}{L_y^2} \epsilon_{x,0}.}
\end{cases}
\end{equation}
Figure~2~(d) in the Letter shows the growth of the geometric average of the emittances after mixing, $(\epsilon_x^*\epsilon_y^*/\epsilon_{x,0}\epsilon_{y,0})^{1/2}$, as a function of $\alpha_x$ and $\alpha_y$,
obtained using Eq.~\eqref{eq:emitt_sat_final} and using the values of $\eta_r$ from Eq.\eqref{eq:fraction_of_resonant_particles}. 

\end{document}